\documentclass[12pt]{article}
\usepackage{epsfig,amsmath,amsfonts,amssymb,amstext,afterpage,psfrag,
}
\usepackage{slashed}
\usepackage[square,comma,sort&compress,numbers]{natbib}
\allowdisplaybreaks
\long\def\symbolfootnote[#1]#2{\begingroup%
\def\thefootnote{\fnsymbol{footnote}}\footnote[#1]{#2}\endgroup}

\def\spose#1{\hbox to 0pt{#1\hss}}
\def\lsim{\mathrel{\spose{\lower 3pt\hbox{$\mathchar"218$}}
 \raise 2.0pt\hbox{$\mathchar"13C$}}}
\def\gsim{\mathrel{\spose{\lower 3pt\hbox{$\mathchar"218$}}
 \raise 2.0pt\hbox{$\mathchar"13E$}}}

\catcode`@=11
\def\@citex[#1]#2{%
  \if@filesw\immediate\write\@auxout{\string\citation{#2}}\fi
  \def\@citea{}\@cite{\@for\@citeb:=#2\do
    {\@citea\def\@citea{,\penalty\@m}\@ifundefined
      {b@\@citeb}{{\bf ?}\@warning
{Citation `\@citeb' on page \thepage \space undefined}}%
      \hbox{\csname b@\@citeb\endcsname}}}{#1}}
\def\citer{\@ifnextchar [{\@tempswatrue\@citexr}{\@tempswafalse\@citexr[]}}
  \def\@citexr[#1]#2{%
    \if@filesw\immediate\write\@auxout{\string\citation{#2}}\fi
    \def\@citea{}\@cite{\@for\@citeb:=#2\do
      {\@citea\def\@citea{--\penalty\@m}\@ifundefined
{b@\@citeb}{{\bf ?}\@warning
{Citation `\@citeb' on page \thepage \space undefined}}%
\hbox{\csname b@\@citeb\endcsname}}}{#1}}

\begin{document}

\begin{titlepage}

\begin{flushright}
{\small
LMU-ASC~10/16\\ 
March 2016\\
}
\end{flushright}

\vspace{0.5cm}
\begin{center}
{\Large\bf \boldmath                                               
Comment on\\ 
\vspace*{0.3cm}
``Analysis of General Power Counting Rules\\
\vspace*{0.3cm}
in Effective Field Theory''
\unboldmath}
\end{center}

\vspace{0.5cm}
\begin{center}
{\sc G. Buchalla, O. Cat\`a, A. Celis and C. Krause} 
\end{center}

\vspace*{0.4cm}

\begin{center}
Ludwig-Maximilians-Universit\"at M\"unchen, Fakult\"at f\"ur Physik,\\
Arnold Sommerfeld Center for Theoretical Physics, 
D--80333 M\"unchen, Germany
\end{center}

\vspace{1.5cm}
\begin{abstract}
\vspace{0.2cm}\noindent
In a recent paper \cite{Gavela:2016bzc} a master formula has been 
presented for the power counting of a general effective field theory.
We first show that this master formula follows immediately
from the concept of chiral dimensions (loop counting),
together with standard dimensional analysis.
Subsequently, \cite{Gavela:2016bzc} has disputed the relevance of 
chiral counting 
for chiral Lagrangians, 
and in particular for the electroweak chiral Lagrangian including a 
light Higgs boson. As an alternative, a power counting
based on `primary dimensions' has been proposed. The difficulties encountered
with this scheme led the authors to suggest that even the leading order
of the electroweak chiral Lagrangian could not be clearly defined.
Here we demonstrate that the concept of primary dimensions is irrelevant
for the organization of chiral Lagrangians. We re-emphasize that
the correct counting is based on chiral dimensions, or the counting of 
loop orders, and show how the problems encountered in 
\cite{Gavela:2016bzc} are resolved.
\end{abstract}

\vfill

\end{titlepage}

\section{Introduction}
\label{sec:intro}

The electroweak chiral Lagrangian including a light Higgs boson
has been developed as a systematic effective field theory (EFT) for the case of
strong dynamics in the Higgs sector in a series of papers
\cite{Buchalla:2012qq,Buchalla:2013rka,Buchalla:2013eza,Buchalla:2014eca},
building on previous work by many authors
\cite{Coleman:1969sm,Callan:1969sn,Weinberg:1978kz,Appelquist:1980vg,Longhitano:1980iz,Manohar:1983md,Gasser:1983yg,Appelquist:1984rr,Cvetic:1988ey,Dobado:1989ax,Dobado:1989ue,Dobado:1989gr,Dobado:1990jy,Dobado:1990zh,Espriu:1991vm,Herrero:1992zq,Herrero:1993nc,Feruglio:1992wf,Urech:1994hd,Nyffeler:1999ap,Knecht:1999ag,Ecker:2000zr,Agashe:2004rs,Giudice:2007fh,Contino:2010mh,Contino:2010rs}.
Phenomenological implications of the formalism, in particular to test anomalous
Higgs couplings at the LHC, 
have been discussed in \cite{Buchalla:2015wfa,Buchalla:2015qju}.

In order to define the electroweak chiral Lagrangian as a systematic
EFT, it is necessary to specify basic assumptions. 
These concern the {\it particle content} below a  
mass gap (assumed to be of the order of TeV), the relevant {\it symmetries}, 
and the {\it power counting}. In \cite{Buchalla:2012qq,Buchalla:2013rka,Buchalla:2013eza,Buchalla:2014eca} the following assumptions were made:
\begin{description}
\item[(i)] 
SM particle content, where (transverse) gauge bosons and fermions
are weakly coupled to the Higgs-sector dynamics. 
\item[(ii)] 
SM gauge symmetries;
conservation of lepton and baryon number; 
conservation {\it at lowest order\/} of custodial symmetry, CP invariance
in the Higgs sector and fermion flavour. 
The latter symmetries are violated at some level, but this 
would only affect terms at subleading order.
Generalizations may in principle be introduced if necessary.
\item[(iii)] 
Power counting by chiral 
dimensions \cite{Urech:1994hd,Nyffeler:1999ap,Knecht:1999ag},
equivalent to a loop expansion \cite{Buchalla:2013eza}, with the simple 
assignment of $0$ for bosons (gauge fields $X_\mu$, Goldstones $\varphi$ and 
Higgs $h$) and $1$ for each derivative, weak coupling (e.g. gauge or Yukawa),
and fermion bilinear:
\begin{equation}\label{chidim}
d_\chi[X_\mu, \varphi, h] = 0\, ,\qquad 
d_\chi[\partial_\mu, g, y, \psi\bar\psi] = 1 
\end{equation}
The total chiral dimension $d_\chi$ of a term in the Lagrangian
determines its loop order $L$ through $d_\chi = 2L+2$.
\end{description}

This picture for the electroweak chiral Lagrangian including
a light Higgs has recently been questioned in \cite{Gavela:2016bzc},
however, without addressing the EFT assumptions in detail
(see also \cite{Alonso:2012px,Alonso:2012pz,Brivio:2013pma}).
In particular, the relevance of chiral dimensions for the
EFT power counting has been disputed, even for the standard case
of pion chiral perturbation theory ($\chi$PT).
In the following we critically examine the results and arguments
presented in \cite{Gavela:2016bzc}. We show that the framework
of the electroweak chiral Lagrangian as outlined above is consistent
and we emphasize that chiral counting is the relevant counting in this case.   
We also take the opportunity to illustrate the workings of chiral dimensions
in a toy scenario of pions, photons and (heavy) leptons.

\section{Master formula for power counting }
\label{sec:power}

In the first part of \cite{Gavela:2016bzc} the derivation of
general power counting rules for effective field theory is reviewed
and a master formula is presented. Here we show how this result
follows from the counting of chiral dimensions, together with standard
dimensional analysis.

We assume an effective theory for light scalars $\phi$,
gauge fields $A$ and fermions $\psi$ at a typical scale $f$ much below a 
cutoff $\Lambda$\footnote{In some models of EWSB the electroweak scale $v$ 
differs from the breaking scale $f$. However, for power-counting 
considerations, both may consistently be taken to be of the same order
$v\sim f\ll \Lambda$.} with gauge, Yukawa and quartic scalar couplings 
$g$, $y$ and $\lambda$. We will work in $d=4$ dimensions. Generalizations to 
arbitrary dimensions are possible but inessential for our discussion.

A generic term in the EFT Lagrangian can then be written as
\begin{equation}\label{lterm}
\partial^{N_p}\, \phi^{N_\phi}\, A^{N_A}\, \psi^{N_\psi}\, 
g^{N_g}\, y^{N_y}\,  \lambda^{N_\lambda}
\end{equation} 
The task of power counting is to determine the (parametric) size
of the coefficient of this term in the Lagrangian.
Using the results of \cite{Buchalla:2013eza}, the loop order $L$ of the
coefficient, or equivalently the power of $1/16\pi^2$, is given by the
chiral dimension $d_\chi\equiv 2L+2$ of the term.   
The canonical dimension $d_c$ of the term and the requirement that
the Lagrangian has dimension 4 then fixes the dependence on the 
dimensionful parameter $f$. Loop factors are thus defined, as usual, 
with respect to the typical
scale of the theory. The coefficient of the term in (\ref{lterm}) then reads
\begin{equation}\label{cterm}
\frac{f^{4-d_c}}{(4\pi)^{d_\chi - 2}}
\end{equation} 
up to a factor of order unity.
The chiral dimension as defined in \cite{Buchalla:2013eza} and the canonical 
dimension are given by
\begin{eqnarray}\label{dcdchi}
d_\chi &=& N_p +\frac{N_\psi}{2} + N_g + N_y + 2 N_\lambda \nonumber\\
d_c &=& N_p + N_\phi + N_A + \frac{3}{2}N_\psi
\end{eqnarray}
Assuming $\Lambda\equiv 4 \pi f$ \cite{Manohar:1983md}, 
the product of the coefficient 
(\ref{cterm}) and the operator (\ref{lterm}) can be written as
\begin{equation}\label{eftmaster}
f^2 \Lambda^2 \left[\frac{\partial}{\Lambda} \right]^{N_p}
\left[\frac{\phi}{f} \right]^{N_\phi}
\left[\frac{A}{f} \right]^{N_A}
\left[\frac{\psi}{f\sqrt{\Lambda}} \right]^{N_\psi}
\left[\frac{g}{4\pi} \right]^{N_g}
\left[\frac{y}{4\pi} \right]^{N_y}
\left[\frac{\lambda}{16\pi^2} \right]^{N_\lambda}
\end{equation}
up to factors of order unity. This result is equivalent to the master formula 
quoted in eq.~(28) of \cite{Gavela:2016bzc}. Eliminating $f=\Lambda/4\pi$, 
(\ref{eftmaster}) takes the form
\begin{equation}\label{eftmaster2}
\frac{\Lambda^4}{16\pi^2} \left[\frac{\partial}{\Lambda} \right]^{N_p}
\left[\frac{4\pi\phi}{\Lambda} \right]^{N_\phi}
\left[\frac{4\pi A}{\Lambda} \right]^{N_A}
\left[\frac{4\pi\psi}{\Lambda^{3/2}} \right]^{N_\psi}
\left[\frac{g}{4\pi} \right]^{N_g}
\left[\frac{y}{4\pi} \right]^{N_y}
\left[\frac{\lambda}{16\pi^2} \right]^{N_\lambda}
\end{equation}
identical to eq.~(22) of \cite{Gavela:2016bzc} and equivalent to the result
already obtained in \cite{Buchalla:2013eza}. 

We emphasize that the concept of chiral dimensions ensures that all the terms 
in the leading-order Lagrangian are homogeneous. The notion of chiral 
number $N_\chi\equiv N_p+N_\psi/2$ used in \cite{Gavela:2016bzc}
is different as the number of couplings is not considered. 

Beyond the case of (\ref{eftmaster2}), the $4\pi$ counting
may be generalized, as discussed in \cite{Gavela:2016bzc}.
This generalisation consists in an independent rescaling of each
of the conserved quantities $g$, $y$, $\lambda$
by an arbitrary factor of $(4\pi)^\nu$, with the power $\nu$ not 
necessarily an integer. Such a rescaling is equivalent to keeping track
of the number of each of these couplings separately. 
It then allows one to expand results in powers of $g$, $y$ or $\lambda$,
in addition to the EFT expansion. It is already clear from
(\ref{eftmaster}) that this can always be done if desired.
Generically, however, the weak couplings $g$, $y$, $\lambda$ 
can be taken to be of ${\cal O}(1)$, and no separate expansion is required.

It should be emphasized that (\ref{eftmaster}) and (\ref{eftmaster2}) encode 
the topological constraints of a consistent power counting for a generic
perturbative theory. However, in addition to this, some dynamical information 
needs to be provided in order to define the full systematics of the expansion.
This systematics is different depending on whether  
the underlying dynamics is weakly or 
strongly coupled.\footnote{In Ref.~\cite{Gavela:2016bzc} the distinction
between weak and strong coupling refers to the size of the couplings 
$g$, $y$, $\lambda$. However, this only means that the theory is inside or 
outside the perturbative regime. What really 
determines the counting is the nature of the underlying dynamics.} 
To define it, the role of the couplings $g$, $y$, $\lambda$, which enter the 
master formula (\ref{eftmaster2}) in a nontrivial way, has to be specified.

As a first example, {\it assume} that all fields are weakly coupled to
the heavy sector, such that a weak coupling can be associated with
every field in the master formula. Eq.~(\ref{eftmaster2}) then becomes
\begin{equation}\label{smeftmaster}
\frac{\Lambda^4}{\kappa^2} \left[\frac{\partial}{\Lambda} \right]^{N_p}
\left[\frac{\kappa\phi}{\Lambda} \right]^{N_\phi}
\left[\frac{\kappa A}{\Lambda} \right]^{N_A}
\left[\frac{\kappa\psi}{\Lambda^{3/2}} \right]^{N_\psi}
\left[\frac{\kappa}{4\pi} \right]^{N'_\kappa}
\end{equation}
In order to simplify the notation we have introduced a generic weak coupling
$\kappa$, which may stand for $g$, $y$ or $\sqrt{\lambda}$, as appropriate.
To obtain (\ref{smeftmaster}) we have simply pulled a factor of $\kappa/4\pi$
in front of every field, and written an overall factor of $(\kappa/4\pi)^{-2}$ 
to keep the canonical normalization of the kinetic terms. The nontrivial 
dynamical assumption here is that the residual number of couplings is
$N'_\kappa\geq 0$. Setting for the moment $N'_\kappa= 0$,
(\ref{smeftmaster}) is then seen to reduce to the standard power counting
by canonical dimensions, where higher-dimensional operators are suppressed
by inverse powers of a large energy-scale $\Lambda$. 
For a given order in $1/\Lambda$,
$N'_\kappa$ counts the number of loop corrections.  

As a second example, consider the chiral perturbation theory of pions
interacting with the photon field (see \cite{Buchalla:2013eza} for 
a more detailed discussion). 
In this case the pions are strongly coupled to the heavy sector and cannot be 
associated with a weak coupling
that would multiply each field. The photon, on the other hand, is still
weakly coupled $\sim e$ and (\ref{eftmaster2}) becomes  
\begin{equation}\label{chptmaster}
\frac{\Lambda^4}{16\pi^2} \left[\frac{\partial}{\Lambda} \right]^{N_p}
\left[\frac{\phi}{f} \right]^{N_\phi}
\left[\frac{A}{f} \right]^{N_A}
\left[\frac{e}{4\pi} \right]^{N_e}
\end{equation}
Here $f$ can be identified with the pion decay constant, related to the 
cutoff $\Lambda\equiv 4\pi f$ through 
a (loop) factor $4\pi$ \cite{Manohar:1983md}.

The EFT described by (\ref{chptmaster}) is valid at energies of order
$f\ll\Lambda$ and the expansion parameter is $f^2/\Lambda^2=1/16\pi^2$,
corresponding to a loop factor.\footnote{The important point is that
this theory is renormalizable only order by order. The loop expansion
guarantees this renormalizability, while an expansion by canonical dimension 
does not.}
With this identification the EFT can be 
viewed as being organized in terms of a loop expansion, equivalent to
a counting of chiral dimensions. As an example, consider the 
operator $e^2\langle U^\dagger QUQ\rangle$, where $Q$ is the quark
electric charge and $U$ the Goldstone matrix. 
This term represents a radiatively induced pion
potential and contains the electromagnetic correction to the pion mass.
According to (\ref{chptmaster}) its coefficient is
$(\Lambda^4/16\pi^2)(e/4\pi)^2=e^2 f^4$, corresponding to a leading-order
coefficient for $e={\cal O}(1)$. An identical result is obtained from
counting the chiral dimension of this operator, which is
$d_\chi=2$ from the factor $e^2$. Of course, the equivalence
of (\ref{chptmaster}) with the counting of chiral dimensions holds in
general, as has been demonstrated above. 

The following conclusions can be drawn:
\begin{itemize}
\item
There are essentially two different expansions that may be used to organize 
a low-energy EFT: The expansion in inverse powers of a high-energy scale and 
the loop expansion. They are governed by the canonical 
dimension and the chiral dimension, respectively, of operators in the 
Lagrangian. In addition, a separate expansion in powers of weak couplings
$\kappa$ can be performed. This expansion is a special case of the
more general scenario which treats $\kappa$ as a quantity of order one.
It is independent of the loop expansion. The above possibilities
of organizing an EFT have been discussed in \cite{Buchalla:2013eza}.
\item
Both (\ref{smeftmaster}) and (\ref{chptmaster}) are obtained as special
cases from the master formula in (\ref{eftmaster2}).
Nevertheless they represent a different organization of the respective
EFT: (\ref{smeftmaster}) by canonical dimensions, (\ref{chptmaster})
by chiral counting. The difference is related to the role of (weak) 
couplings that needs to be specified in addition to (\ref{eftmaster2}).
\item
The power counting of chiral perturbation theory and its extensions, 
e.g. when coupled to photons, is governed by chiral dimensions 
(see also Sec. \ref{sec:toymod} below). The consistency of the framework 
can be proven by comparison with the explicit determination 
of the one-loop effective action~\cite{Knecht:1999ag}. 
\item
The counting of weak couplings due to their chiral dimension
may seem unfamiliar. However, it can be relevant even in the context
of the usual standard-model operators of dimension 6 
\cite{Buchmuller:1985jz,Grzadkowski:2010es}.
Consider for example the Higgs-gluon operator 
$H^\dagger H G_{\mu\nu}G^{\mu\nu}$.
Dimensional counting alone only implies that the coefficient is 
$\sim 1/\Lambda^2$. Assuming now that all fields are weakly coupled
implies that the operator comes with a factor of $y^2 g^2_s$ 
(or equivalent combinations in terms of chiral dimensions).
The master formula (\ref{eftmaster2}) then yields
 \begin{equation}\label{hhgg}
\frac{1}{16\pi^2}\frac{1}{\Lambda^2} y^2 g^2_s H^\dagger H G_{\mu\nu}G^{\mu\nu}
\end{equation}
Therefore, under the assumption of weak coupling to the heavy physics,
the coeffcient is suppressed by an additional loop factor, as has been
discussed in \cite{Grzadkowski:2010es,Arzt:1994gp}.
Since $1/(16\pi^2\Lambda^2)=1/(16\pi^2)^2f^2$, 
this double suppression amounts (effectively) to a suppression by
two loop orders (even though the operator may be induced by a one-loop
diagram with internal particles of 
mass $\Lambda$ \cite{Huo:2015exa,Dolan:2016eki}).
Note that this double suppression is correctly reproduced by
the chiral dimension of $y^2 g^2_s H^\dagger H G_{\mu\nu}G^{\mu\nu}$,
which amounts to $d_\chi=2L+2=6$. 
\end{itemize}

The electroweak chiral Lagrangian with a light Higgs implies strongly-coupled 
dynamics and therefore is organized through a loop expansion, with a power 
counting in terms of chiral dimensions as defined in 
eq.~(\ref{dcdchi}) \cite{Buchalla:2013eza}.
A priori, the electroweak vev $v$ and the scale of the strong dynamics $f$
need not be distinguished for the purpose of power counting.
This is equivalent to treating $\xi=v^2/f^2$ as a quantity of order one.
The EFT is then very similar to chiral perturbation theory with pions and
photons. However, a further expansion in powers of $\xi$ can still be
performed. This results in an EFT with a double expansion in both loop order 
and in canonical dimensions, as explained in \cite{Buchalla:2014eca}.

\section{Cross sections}
\label{sec:cross}

Section IV of \cite{Gavela:2016bzc} considers the power counting
of cross sections. 
For the case of pion scattering in chiral perturbation theory, the following 
counting rules are quoted
\begin{equation}\label{sigmak2}
\sigma_k(\varphi\varphi\to \varphi\varphi) \sim
\frac{\pi(4\pi)^2}{E^2}\frac{E^4}{\Lambda^4}
\end{equation}
\begin{equation}\label{sigma42}
\sigma_4(\varphi\varphi\to \varphi\varphi) \sim
\frac{\pi(4\pi)^2}{E^2}\frac{E^8}{\Lambda^8}
\end{equation}
\begin{equation}\label{sigmak4}
\sigma_k(\varphi\varphi\to 4\varphi) \sim
\frac{\pi(4\pi)^2}{E^2}\frac{E^8}{\Lambda^8}
\end{equation}
The amplitudes for the scattering of pions $\varphi$ are evaluated 
at tree level with interactions from the Lagrangian of ${\cal O}(p^2)$ 
($\sigma_k$) or ${\cal O}(p^4)$ ($\sigma_4$). Factors of energy or momentum
are denoted by $E$, and $\Lambda=4\pi f$ is the EFT cutoff.

We agree with the results in (\ref{sigmak2}) -- (\ref{sigmak4}), but would 
like to comment on the interpretation given in \cite{Gavela:2016bzc}.
In that paper, a comparison is made between 
(\ref{sigma42}) and (\ref{sigmak4}). The fact that their scaling
with $\Lambda$ is identical, despite the different chiral counting
of the underlying interaction, is taken as an indication against
the relevance of chiral counting in this context.
We disagree with such a conclusion.
First, (\ref{sigma42}) and (\ref{sigmak4}) refer to different
processes. Since the number of final-state particles is different,
the phase-space factors are not the same. It is then unclear
what should be inferred from such a comparison.
More meaningful is the comparison between (\ref{sigmak2}) and
(\ref{sigma42}), which represent different contributions to
the same process. However, the next-to-leading correction $\Delta\sigma_{k,4}$
to the leading-order cross section (\ref{sigmak2}) is not given
by (\ref{sigma42}), but by the interference of the leading-order
amplitude from ${\cal L}_2$ and the next-to-leading order term from
${\cal L}_4$, leading to
\begin{equation}\label{sigmak242}
\Delta\sigma_{k,4}(\varphi\varphi\to \varphi\varphi) \sim
\frac{\pi(4\pi)^2}{E^2}\frac{E^6}{\Lambda^6}
\end{equation}
The correction is of order $E^2/\Lambda^2\equiv E^2/16\pi^2 f^2$ relative 
to (\ref{sigmak2}). This is exactly of the size expected from chiral counting
where the next-to-leading terms are suppressed by a loop factor.
Since $\Lambda=4\pi f$, the presence of inverse powers of $\Lambda$
in the formulas above is in agreement with the loop expansion and the
chiral counting. The statement in the abstract of \cite{Gavela:2016bzc}
that ``the size of cross sections is controlled by the $\Lambda$ power
counting of EFT, not by chiral counting, even for chiral perturbation
theory'' is therefore not justified. 

The latter conclusion can be made even more explicit by noting
that the formulas in (\ref{sigmak2}) -- (\ref{sigmak4}) can
indeed be obtained through the counting of chiral dimensions.
For this purpose we employ the optical theorem, also considered 
in \cite{Gavela:2016bzc}, which we write schematically as
\begin{equation}\label{optical}
\sigma(i\to f)\sim\frac{1}{E^2}\, {\rm Im}{\cal M}(i\to f \to i)
\end{equation}
The derivation of (\ref{sigmak2}) may suffice as an example.
In this case, the forward amplitude ${\cal M}$, with $i=f=\varphi\varphi$,
is a one-loop amplitude with two vertices from the ${\cal O}(p^2)$
Lagrangian, and thus an ${\cal O}(p^4)$ term, according to chiral counting
(chiral counting reduces to momentum counting in pion chiral perturbation
theory). We then have ($p\sim E$)
\begin{equation}\label{mme4}  
{\cal M}(i\to f \to i)\sim E^4
\end{equation}
Since ${\cal M}$ is dimensionless, the energy dependence has to be $(E/f)^4$.
The compensating scale can only be $f$, not $\Lambda$, as the
loop amplitude in the low-energy EFT depends only on the physics in the IR,
not in the UV. In addition, a chiral dimension of $d_\chi\equiv 2L +2=4$ 
implies a loop order of $L=1$, and thus a factor of $1/16\pi^2$ for the
coefficient in (\ref{mme4}). Therefore, 
\begin{equation}\label{mmef4}  
{\cal M}(i\to f \to i)\sim \frac{E^4}{f^4} \frac{1}{16\pi^2}
\end{equation}
Inserting (\ref{mmef4}) in (\ref{optical}) and using that the imaginary part
yields a factor of $\pi$, one finds
\begin{equation}\label{sigk2chi}
\sigma_k(\varphi\varphi\to \varphi\varphi) \sim
\frac{\pi}{E^2}\frac{E^4}{f^4} \frac{1}{16\pi^2}
\end{equation}
which is equivalent to (\ref{sigmak2}).
The remaining formulas, and similar ones, can be derived along the
same lines.

We finally comment on the counting of $E/f$ factors in chiral
perturbation theory. In general, these factors can be taken to
be of order unity, corresponding to $E\sim f$, where $f\ll\Lambda$ is 
the energy scale at which the EFT is valid. Such a counting is fully
consistent with the loop expansion and therefore with the
order-by-order renormalization of the chiral Lagrangian.
If $E$ is assumed to be numerically larger than $f$,
$E^2/f^2$ terms may be considered to be enhanced with respect to
other terms of the same chiral order such as $m^2_\pi/f^2$.
Any result based on chiral counting can then be further approximated
exploiting this enhancement. However, this does not invalidate
the (more general) results obtained under the assumption $E\sim f$.
In any case, the enhancement of $E$ is limited by the requirement
that $E\ll\Lambda$.

\section{Electroweak chiral Lagrangian}
\label{sec:ewchil}

The electroweak chiral Lagrangian including a light Higgs, referred to as 
HEFT in \cite{Gavela:2016bzc}, is discussed in section V of that paper.
We disagree with several of the assertions made there. In the following we will
address the most relevant points.

\vspace*{0.5cm}

\begin{itemize}
\item 
The authors of \cite{Gavela:2016bzc} introduce the concept
of {\it primary dimension} $d_p$ as an ordering principle
for chiral Lagrangians. The primary dimension of a quantity $R$
is defined as the canonical dimension of the first nonvanishing term
in an expansion of $R$ in powers of field variables.
For instance, the Goldstone matrix $U=\exp(2i\Pi^a T^a/f)$,
$T=U\sigma_3 U^\dagger$ and $V_\mu = (\partial_\mu U)U^\dagger$  have
$d_p$ equal to 0, 0 and 2, respectively. 

We argue that the primary dimension $d_p$ is irrelevant for 
the power counting of chiral Lagrangians.
This follows immediately from a simple counterexample.
Consider the operators
\begin{equation}\label{kinspar}
-\frac{1}{4}B_{\mu\nu} B^{\mu\nu},\qquad
g'g\langle U T_3 B_{\mu\nu} U^\dagger W^{\mu\nu}\rangle
\end{equation}
of the (Higgsless) electroweak chiral Lagrangian, where the first
is the $B$-field kinetic term, and the second corresponds to the
electroweak $S$-parameter. These operators enter the chiral Lagrangian
at leading and next-to-leading order, 
respectively \cite{Appelquist:1980vg,Longhitano:1980iz}.
The concept of primary dimension is unable to reproduce this crucial 
distinction since both terms are assigned $d_p=4$. Consequently,
counting primary dimensions does not lead to the correct power counting
for the electroweak chiral Lagrangian (with or without light Higgs).

Table II of \cite{Gavela:2016bzc} lists operators of the
electroweak chiral Lagrangian together with their primary dimension
$d_p$, canonical dimension $d$ and $N_\chi$ (the chiral dimension of a term
up to the contribution of any weak couplings).
Like $d_p$, neither $d$ nor $N_\chi$ can distinguish the order of the
terms in (\ref{kinspar}), yielding $d=4$ and $N_\chi=2$ for both. 

The problem is resolved with the rules of chiral counting, which
imply a chiral dimension of $d_\chi=2$ for the gauge kinetic term,
and $d_\chi=4$ for the $S$-parameter term, indicating the correct
order of the terms in the chiral Lagrangian. As this
example shows, keeping track of the chiral dimensions of the weak couplings 
is essential to get the right counting.
\item
We disagree with the assertion made in the first column of page~11
in \cite{Gavela:2016bzc} that $U=\exp(2i\Pi^a T^a/f)$, $V_\mu$ or $T$
contain ``hidden factors of $\Lambda$'', with $\Lambda=4\pi f$ the
cutoff of the chiral Lagrangian.
This view contradicts basic EFT principles, according to which the physics
at high energy is fully contained in operator coefficients, whereas
EFT field variables, such as $U$, encode the low-energy (IR)
dynamics and are independent of the UV. 
\item
In an attempt to construct the leading-order of the 
electroweak chiral EFT including a light Higgs,
eq. (62) of \cite{Gavela:2016bzc} defines a Lagrangian
${\cal L}^{d_p\leq 4}$ that collects the terms with $d_p\leq 4$.
The resulting expression is incorrect for several reasons. 
\begin{itemize}
\item[(i)]
Unlike the LO gauge-kinetic terms $X^2_{\mu\nu}$, the operators 
$g^2 X_{\mu\nu}^2 h$ arise only at NLO. The latter operators could only
appear at leading order if the gauge fields were strongly coupled to the
heavy sector, which is not the case and would in fact be inconsistent.  
The expression ${\cal L}^{d_p\leq 4}$ fails to account for this important
feature, listing the above interaction terms at the same (leading) order.
\item[(ii)]
The Goldstone kinetic term $-(f^2/4)\langle V_\mu V^\mu\rangle$
(which comes with an incorrect sign in \cite{Gavela:2016bzc})
implies a $W$-boson mass of $M_W=gf/2$, rather than the correct 
value $M_W=gv/2$. The problem arises since the presentation
in \cite{Gavela:2016bzc} does not carefully distinguish the
electroweak vev $v$ from the new physics scale $f$.
\item[(iii)]
The Higgs potential and the Higgs mass scale as $\Lambda^2$,
while the size of the coefficient is left unspecified.
Chiral counting dictates that the chiral dimension of this
coefficient has to be taken into account. Consistency requires
that the coefficient carries a chiral dimension of 2, implying
an electroweak-scale mass for the light Higgs. One explicit realization
of this occurs in composite-Higgs models where the Higgs potential
is generated radiatively. It then comes with two powers of weak couplings
($g^2$, $y^2$) and a corresponding loop suppression of the
scale $\Lambda^2$. We note that there is no ambiguity here: The
chiral dimension of the coefficient cannot be ignored, otherwise
the Higgs mass would be of the size of the cutoff $\Lambda$ and it
would be inconsistent to include the Higgs as a field in the EFT. 
The assumptions spelled out in Sec. \ref{sec:intro} reflect
a similar consistency requirement for the gauge-boson and fermion sectors.
\end{itemize} 
\item
The comments following eq.~(65) of \cite{Gavela:2016bzc} suggest
to associate the operators 
\begin{equation}\label{ChPT4}
e^2\langle F^2_{R\mu\nu} + F^2_{L\mu\nu}\rangle, \qquad
e^2\langle U^\dagger F_{R\mu\nu} U F^{\mu\nu}_L\rangle
\end{equation}
of chiral perturbation theory with
the leading-order photon kinetic term in the power counting.
This contradicts the well-known fact that these operators enter the 
Lagrangian as counterterms at next-to-leading order \cite{Urech:1994hd}. 
Again, the correct classification follows immediately from the
chiral dimension $d_\chi=4$ of these terms, whereas the primary dimension
fails to capture the difference between~(\ref{ChPT4}) and the leading-order 
kinetic term ($d_p=4$ for both). 
\item
At the end of Sec.~V in \cite{Gavela:2016bzc} various attempts to define 
a leading-order Lagrangian are considered, in particular
based on the criteria $N_\chi\leq 2$ or $d_p\leq 4$, or
else using a simultaneous counting in both $N_\chi$ and $d_p$. 
As discussed above, none of these options is valid.
\end{itemize}

\section{A toy model}
\label{sec:toymod}

To illustrate how the power counting based on chiral dimensions
works in a setting where Goldstone dynamics is coupled to gauge fields
and fermions, the following example may be considered.
Take the pions of QCD, together with a light lepton $\psi$ of mass $m$
(e.g. of order $100\,{\rm MeV}$)
and a heavy lepton $\Psi$ of mass $M$, gauged in the usual way under
the electromagnetic $U(1)$. A theory of this type, with quarks that are 
either electrically charged or neutral, has also been discussed 
in \cite{Gavela:2016bzc}. Here we assume the realistic case of charged
quarks as the difference to neutral quarks is immaterial for our discussion.
The pion dynamics has a cutoff $\Lambda=4\pi f$ of order $1\,{\rm GeV}$.
The QED coupling is weak, $e={\cal O}(1)$. 
Whereas in nature $e\approx 0.3$, it will be illuminating to also
imagine a toy scenario where $e=1$ numerically. Perturbativity in QED would 
still hold,\footnote{Electromagnetism is perturbative not because $e$ is 
numerically small, but because of the loop expansion, with $e={\cal O}(1)$.} 
but a further expansion in $e$ could no longer be performed.

In addition to $\Lambda$ the theory contains another large scale $M$.
It is true, as mentioned in \cite{Gavela:2016bzc}, that $\Lambda$ and $M$
are in principle unrelated. Three possibilities may then be distinguished:
$M\ll\Lambda$, $M\sim\Lambda$, and $M\gg\Lambda$.

In the first case, the lepton $\Psi$ cannot be integrated out and
will simply remain an explicit degree of freedom of the low-energy EFT.

In the second case, the scales $M$ and $\Lambda$ can be identified for
the purpose of power counting. While their physical origin might be very 
different, the ratio $\Lambda/M$ is a number of order unity.
When the physics at $M\sim\Lambda$ is integrated out, $\Lambda/M$ will
be encoded in the ${\cal O}(1)$ coefficients of the EFT Lagrangian.
This is the standard case for a bottom-up construction of the EFT, in which 
the details of the physics at the cutoff are unknown.

In the third case, the effects of $\Psi$ in the EFT coefficients would be
suppressed by powers of $\Lambda/M\ll 1$. Some coefficients might tend to 
zero or other simplifications might occur. However, no information will
be lost in the usual formulation of the EFT with general coefficients.

The issue of a possible hierarchy among two (or more) different heavy
mass scales is clearly not restricted to the case of a chiral Lagrangian.
It could also arise for weakly-coupled theories where the low-energy EFT
is organized by canonical dimensions. Suppose such an EFT describes
physics at a scale $v\ll\Lambda_1\ll\Lambda_2$. The essential feature is
the mass gap between $v$ and $\Lambda_1$. A general expansion in powers of
$1/\Lambda_1$ will provide the most general set of higher-dimensional
operators, where ${\cal O}(1)$ coefficients encode the physics at 
energies $\sim\Lambda_1$ and above, irrespective of its details
and of the presence of a further scale $\Lambda_2$. What matters
is the particle content, the symmetries, and the power counting.

The authors of \cite{Gavela:2016bzc} insist that when a theory contains 
both strongly-coupled and weakly-coupled dynamics, there is no unified 
counting. If this were true, it would imply that interactions of pions with 
photons are beyond an EFT description. The covariant derivative 
$D_{\mu}U=\partial_{\mu}U +ieA_{\mu}[Q,U]$ unavoidably 
mixes the strongly-coupled and the weakly-coupled sector. The combined 
dynamics can still be consistently organized in terms of a loop expansion.
As a result, we disagree with the statement in Sec.~V of 
\cite{Gavela:2016bzc} that ``it is not helpful to force both sectors into a 
unified counting with a single expansion parameter''. 
On the contrary, a well-defined counting is an essential ingredient 
of any bottom-up EFT.

One can use the previous example to illustrate how the power-counting formula 
works depending on whether the dynamics is strongly or weakly coupled. 
At scales $f\ll\Lambda$, $M$, the heavy fermion $\Psi$ can be integrated out. 
The EFT Lagrangian will contain (among others) the terms
\begin{eqnarray}\label{toymod}
{\cal L}_{eff} &=& \frac{f^2}{4}\langle D_\mu U^\dagger D^\mu U\rangle
+\bar\psi (i\!\not\!\! D - m)\psi
-\frac{1}{4}F_{\mu\nu} F^{\mu\nu} +\ldots \nonumber\\
&& + a_1  e^4 (F_{\mu\nu} F^{\mu\nu})^2 
+ a_2 e^4 F_{\mu\nu} F^{\nu\sigma} F_{\sigma\rho} F^{\rho\mu}
+\ldots 
\end{eqnarray}
The first line contains leading-order terms, scaling as $f^4$
at energies of order $f$ (the scale where the EFT is valid).
The second line displays the well-known operators from the 
Euler-Heisenberg (EH) Lagrangian (see e.g. \cite{Pich:1995bw}), arising  
once the heavy lepton $\Psi$ has been integrated out. 
These $(F_{\mu\nu})^4$ operators scale as $f^8$ from dimensional analysis. 

Consider first a theory without pions. The EH operators are then catalogued 
as $d=8$ with one loop suppression, where
\begin{equation}\label{a12eh}
a_1=-\frac{1}{36(16\pi^2) M^4}\, ,\qquad\quad a_2=\frac{7}{90(16\pi^2) M^4} 
\end{equation}
One can easily check that both the right 
dimensional scaling and the number of loop suppressions follow 
straightforwardly from eq.~(\ref{smeftmaster}). The coefficients of the 
local operators can be determined exactly 
because we know the UV physics that has been integrated out. 
In other words, this is an example of a top-down EFT. 

When pions are present and $M\sim\Lambda=4\pi f$ one should use instead 
eq.~(\ref{chptmaster}) as the power-counting formula. Additional operators 
appear, mixing the dynamics of pions and photons (and light fermions $\psi$). 
According to our previous discussion, this EFT should now be organized with 
chiral dimensions. This might seem surprising 
because the operators shown in the second line of~(\ref{toymod}) do not 
contain explicit pion fields. However, we note that those operators have 
now an additional contribution coming from 3-loop pion exchange diagrams,
which require $a_{1,2}$ as counterterms. 
This is precisely what eq.~(\ref{chptmaster}) indicates: counting 
$M\sim\Lambda=4\pi f$, the operators $e^4(F_{\mu\nu})^4$ are suppressed 
by three powers of the loop factor $1/(16\pi^2)^3$ with respect to the leading 
order. 
Adding the pions has therefore changed the dynamics 
and the expansion of the EFT. 
The EFT is now a bottom-up one and the coefficient of $(F_{\mu\nu})^4$ 
operators contains hadronic dynamics parametrically of the same size as the 
contribution of $\Psi$.

This result is readily obtained using the counting of chiral dimensions.
We are given the operator $e^4 (F_{\mu\nu})^4$ within the EFT of the model
above. The problem is to find the magnitude of the coefficient.
First, we note that $(F_{\mu\nu})^4$ has to come with at least four powers of
the coupling $e$. This is because we know that each photon field is weakly 
coupled $\sim e$ to the heavy sector that has been integrated out.
The chiral dimension of this operator is 8 (4 derivatives and 4 couplings).
$d_\chi\equiv 2 L + 2 =8$ implies loop order $L=3$. This gives
an estimate for the coefficient of $1/(16\pi^2)^3$, up to factors of
order unity and $1/f^4$ from dimensional analysis, 
in agreement with the explicit result in (\ref{toymod}), (\ref{a12eh}). 

It is important to realize the conceptual difference between
a bottom-up and a top-down construction of an EFT.
In the top-down case the physics in the UV is known and the
EFT is constructed as its low-energy approximation.
Clearly, all the details of the EFT coefficients are then known
explicitly, as seen for instance in (\ref{a12eh}).
The situation is different for a bottom-up EFT, where only
the particle content, the symmetries and a power counting are
specified, but the detailed dynamics in the UV is unknown.
This is the case of the electroweak chiral Lagrangian. 

For the toy scenario discussed here, the bottom-up perspective
would be the construction of the EFT from the pions, the photon
and the light lepton. Nothing would be known about the existence
of the heavy lepton $\Psi$. Nevertheless, the operators $e^4 (F_{\mu\nu})^4$
would be written at 3-loop order in the EFT. This would be consistent 
with their appearance in the concrete case of (\ref{a12eh}), or with
any similar case where $\Psi$ would be substituted by other
heavy degrees of freedom.   

\begin{figure*}[t]
\begin{center}
\includegraphics[width=12cm]{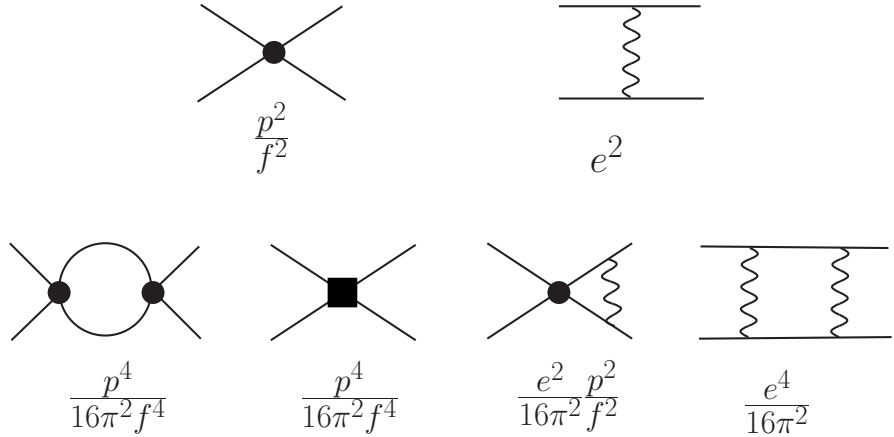}
\end{center}
\caption{\it Sample diagrams for $\pi^+\pi^-\to\pi^+\pi^-$ scattering
in chiral perturbation theory coupled to photons. The first line displays
leading-order, the second line next-to-leading order contributions
in the chiral counting.}
\label{fig:toymod}
\end{figure*}

It is also instructive to see how the EFT counting based on
chiral dimensions works for pion scattering within
the toy model above (where $e=1$ may be chosen).
The model corresponds to pion chiral perturbation theory coupled to
photons and is conceptually similar to the electroweak chiral
Lagrangian. An interesting aspect here is the interplay of
the pions, arising from a strongly coupled sector, with the
weakly coupled photons. This situation can still be treated
by a loop expansion, which is automatically consistent with
the renormalization of the EFT  \cite{Urech:1994hd}.
The resulting systematics is illustrated for $\pi^+\pi^-\to\pi^+\pi^-$ 
scattering at leading and next-to-leading order in Fig. \ref{fig:toymod}.

The first line shows leading-order amplitudes, which are
of order unity since $p\sim f$ and $e={\cal O}(1)$.
They have chiral dimension $d_\chi=2$.
The loop diagrams and local counterterms in the second line
are suppressed by a loop factor. All of them consistently
carry a chiral dimension of $d_\chi=4$.
This scheme provides a consistent counting, irrespective of whether
the interactions come from pion dynamics or from photon exchange.
In particular, the two leading-order contributions are allowed
to be numerically of similar size.
This does not preclude, however, the possibility to make
further approximations if e.g. $p^2/f^2$ is numerically larger
than $e^2$, either because $e\ll 1$, or because $p$ is somewhat
larger than $f$. Indeed, a possible approximation, often considered
in practice, would be to put $e\to 0$, in which case pure 
chiral perturbation theory would be recovered.
Yet the picture in Fig. \ref{fig:toymod} based on
chiral counting remains a consistent and general starting point. 

We remark that the $SU(2)$ gauge coupling $g\approx 0.6$ in the 
electroweak EFT is numerically not far from 1. For electroweak-scale
processes such as Higgs decays typical momentum factors
$\sim p/v$ and $g$ are of comparable order, even numerically.
It is in particular not necessary, and would in fact be inconvenient,
to expand in powers of $g$ or $M_W/v\sim g$. 
Chiral counting takes this feature automatically into account.

\section{Conclusions}
\label{sec:concl}

We have shown that the topological master formulas derived 
in~\cite{Gavela:2016bzc} for generic effective field theories are equivalent 
to the ones already discussed by some of us in previous 
papers~\cite{Buchalla:2012qq,Buchalla:2013rka,Buchalla:2013eza}. 
In those papers it was already emphasized that a topological analysis 
amounts to a unique assignment of chiral dimensions to fields and couplings. 
However, it is important to realize that topological master formulas are 
not power-counting formulas by themselves: one still needs to 
specify the nature of the underlying dynamics. 
As a result, the topological master formulas can lead to expansions in 
canonical dimensions or chiral dimensions, depending on whether the 
underlying dynamics is weakly or strongly coupled. In~\cite{Gavela:2016bzc} 
the choice of dynamics is not carefully spelled out and the concept of 
chiral dimensions incorrectly implemented, since couplings are not given 
chiral dimensions. This leads the authors of~\cite{Gavela:2016bzc} to cast 
doubts on the role of chiral dimensions in scenarios with strongly-coupled 
dynamics and eventually prompts them to introduce the notion of primary 
dimensions. In this comment we have shown that primary 
dimensions lead to serious inconsistencies and cannot be a valid organizing 
criterion for an EFT expansion. Instead, all the inconsistencies are 
dispelled once chiral dimensions are correctly implemented. 
We have illustrated the workings and the usefulness of chiral dimensions 
with various examples including the loop-suppression of certain
dimension-6 operators in the SM, chiral perturbation theory with pions and 
photons, the Euler-Heisenberg Lagrangian, and the generic loop counting for 
amplitudes, cross sections and phase-space factors. 
In particular, 
and contrary to what is concluded in~\cite{Gavela:2016bzc}, chiral dimensions 
define an unambiguous way how to systematically build an 
electroweak chiral EFT with a light Higgs.

\section*{Acknowledgements}

C.K. acknowledges useful discussions with Zhengkang Zhang.
This work was performed 
in the context of the ERC Advanced Grant project `FLAVOUR' (267104) and was 
supported in part by the DFG cluster of excellence `Origin and Structure 
of the Universe' and DFG grant BU 1391/2-1. A.C. is supported
by a Research Fellowship of the Alexander von Humboldt Foundation.


\end{document}